\documentstyle[12pt]{article}
\makeatletter

\newdimen\LENB \newdimen\LENW \newdimen\THI 
\newdimen\LENWH \newdimen\LENTOT \newcount\N 
\def\vbrknlnele#1#2#3{
  \LENB=#1pt \LENW=#2pt \THI=#3pt
  \LENWH=\LENW \divide\LENWH by 2
  \LENTOT=\LENB \advance\LENTOT by \LENW
  \vbox to \LENTOT{
    \vbox to \LENWH{}
    \nointerlineskip
    \vbox to \LENB{\hbox to \THI{\vrule width \THI height \LENB}}
    \nointerlineskip
    \vbox to \LENWH{}
  }}

\def\vbrknln#1{
  \N=#1
  \vcenter{
    \vbox{
      \loop\ifnum\N>0
        \vbox to 4pt{\vbrknlnele{2}{2}{0.1}}
        \nointerlineskip
        \advance\N by -1
      \repeat
  }}}

\def\vbl#1{\hskip-5pt \vbrknln{#1} \hskip-5pt}

\def\hbrknlnele#1#2#3{
  \LENB=#1pt \LENW=#2pt \THI=#3pt
  \LENTOT=\LENB \advance\LENTOT by \LENW
  \vcenter{
    \vbox to \THI{
      \hbox to \LENTOT{
        \hfil
        \vrule width \LENB height \THI
        \hfil}
  }}}

\def\hblele{\hbrknlnele{2}{2.2}{0.1}}

\def\hblfil{\cleaders\hbox{$ \m@th \mkern1mu \hblele \mkern1mu
$}\hfill}

\def\eqnarray{%
\stepcounter{equation}%
\let\@currentlabel=\theequation
\global\@eqnswtrue
\global\@eqcnt\z@
\tabskip\@centering
\let\\=\@eqncr
$$\halign to \displaywidth\bgroup\@eqnsel\hskip\@centering
$\displaystyle\tabskip\z@{##}$&\global\@eqcnt\@ne
\hfil$\displaystyle{{}##{}}$\hfil
&\global\@eqcnt\tw@$\displaystyle\tabskip\z@{##}$\hfil
\tabskip\@centering&\llap{##}\tabskip\z@\cr}

\def\@cite#1#2{\unskip\nobreak\relax
     \def\@tempa{$\m@th{\hbox{[#1]}}$}%
    \futurelet\@tempc\@citexx}

\def\@citexx{\ifx.\let\@tempd=\@citepunct \@tempc\else
    \ifx,\let\@tempd=\@citepunct \@tempc\else
    \let\@tempd=\@tempa\fi\fi\@tempd}
\def\@citepunct{\@tempc\edef\@sf{\spacefactor=\the\spacefactor\relax}\@tempa
    \@sf\@gobble}

\def\citenum#1{{\def\@cite##1##2{##1}\cite{#1}}}
\def\citea#1{\@cite{#1}{}}

\newcount\@tempcntc
\def\@citex[#1]#2{\if@filesw\immediate\write\@auxout{\string\citation{#2}}\fi
  \@tempcnta\z@\@tempcntb\m@ne\def\@citea{}\@cite{\@for\@citeb:=#2\do
    {\@ifundefined
       {b@\@citeb}{\@citeo\@tempcntb\m@ne\@citea\def\@citea{,}{\bf ?}\@warning
       {Citation `\@citeb' on page \thepage \space undefined}}%
    {\setbox\z@\hbox{\global\@tempcntc0\csname b@\@citeb\endcsname\relax}%
     \ifnum\@tempcntc=\z@ \@citeo\@tempcntb\m@ne
       \@citea\def\@citea{,}\hbox{\csname b@\@citeb\endcsname}%
     \else
      \advance\@tempcntb\@ne
      \ifnum\@tempcntb=\@tempcntc
      \else\advance\@tempcntb\m@ne\@citeo
      \@tempcnta\@tempcntc\@tempcntb\@tempcntc\fi\fi}}\@citeo}{#1}}
\def\@citeo{\ifnum\@tempcnta>\@tempcntb\else\@citea\def\@citea{,}%
  \ifnum\@tempcnta=\@tempcntb\the\@tempcnta\else
   {\advance\@tempcnta\@ne\ifnum\@tempcnta=\@tempcntb \else \def\@citea{--}\fi
    \advance\@tempcnta\m@ne\the\@tempcnta\@citea\the\@tempcntb}\fi\fi}

\makeatother
\def\romanno#1{\uppercase\expandafter{\romannumeral#1}}
\def\ol{\overline}

\newtheorem{th}{{\sc Theorem}}[section]
\newtheorem{prop}[th]{{\sc Proposition}}

\newtheorem{lem}[th]{{\sc Lemma}}
\newtheorem{rem}[th]{{\sc Remark}}
\newcommand{\wh}{\widehat}
\newcommand{\qed}{\hbox{\rule[-2pt]{3pt}{6pt}}}
\begin{document}

\begin{titlepage}

\begin{center}

\begin{Large}
{\bf Rational Solutions for the Discrete}\\[2mm]
{\bf Painlev\'e II Equation}\\[3mm]
\end{Large}

\vspace{30pt}

\begin{normalsize}
{\sc Kenji Kajiwara}, {\sc Kazushi Yamamoto}\\[2mm]
{\it Department of Electrical Engineering,
Doshisha University,}\\
{\it Tanabe, Kyoto 610-03, Japan}\\
{and}\\[2mm]
 {\sc Yasuhiro Ohta}\\[2mm]
{\it Department of Applied Mathematics, Faculty of Engineering,}\\
{\it Hiroshima University, }\\
{\it 1-4-1 Kagamiyama, Higashi-Hiroshima 739, Japan}\\
\end{normalsize}
\end{center}
\vspace{30pt}
\begin{abstract}
The rational solutions for the discrete Painlev\'e II equation
are constructed based on the bilinear formalism. It is shown that they
are expressed by the determinant 
whose entries are given by the Laguerre polynomials.
Continuous limit to the Devisme polynomial representation of the
rational solutions for the Painlev\'e II equation is also discussed.
\end{abstract}
\end{titlepage}

\addtolength{\baselineskip}{.3\baselineskip}

\section{Introduction}
Nonlinear integrable discrete systems are now attracting
much attention. Among them, the discrete Painlev\'e equations are
expected to be the most fundamental ones from the analogy
of the continuous cases. Closer studies are now
revealing rich mathematical structures behind them, 
such as existence of Lax Pair, B\"acklund transformation,
singularity confinement property, and so on\cite{dP}.

As for solutions, it is known that the solutions of the Painlev\'e equations
are transcendental in general, but they admit two classes of ``classical solutions'',
namely, special function type solutions and algebraic solutions. It is also
shown that special function type solutions are expressed by determinants
whose entries are given by special functions.
For example,  the special function type solutions for 
the Painlev\'e II equation(P$_{\rm II}$),
\begin{equation}
\frac{d^2}{dt^2}w=2w^3-2tw+\alpha\ ,\label{p2}
\end{equation}
where $\alpha$ is a parameter, are expressed by the determinant
whose entries are given by the Airy function and its derivatives\cite{Okamoto}.

It is natural to expect that the discrete Painlev\'e equations also
admit the special function type solutions with good structure.
Several cases has been studied 
and it is shown that the discrete Painlev\'e equations
admit the particular solutions expressed by the determinants
whose entries are given by the functions which are regarded as 
the discrete analogue of the special functions\cite{dP2,dP3,dP1,alt-dP2}. 

How about the algebraic solutions? 
Recently, it has been shown that the rational solutions of P$_{\rm II}$ 
admit the determinant representation whose entries are given by the
Devisme polynomials\cite{P2}. Thus we also expect that the discrete Painlev\'e
equations admit algebraic solutions which are expressed by
determinants whose entries are given by discrete analogue of some ``classical
object''.

In this article, we discuss the rational solutions for
the discrete Painlev\'e II equation(dP$_{\rm II}$),
\begin{equation}
X(n+1) + X(n-1) = \frac{(an+b)X(n)+c}{1-X(n)^2}\ ,
\end{equation}
where $a$, $b$ and $c$ are parameters.
In ref.\cite{dP2_rational},
a sequence of the rational solutions for dP$_{\rm II}$ has been 
constructed by using the B\"acklund transformation.
However, its determinant representaion has not been studied well.
We present a discrete analogue of the Devisme polynomial representation
for the rational solutions of P$_{\rm II}$ and show that
the entries are given by the Laguerre polynomials.
We also show that they reduce to the rational solutions of
P$_{\rm II}$ in the continuous limit with the suitable
parametrization.

 \section{Main Result}
In this section, we state our main result.

Let $L_k^{(n)}(x)$ be the Laguerre polynomial defined by
\begin{equation}
\sum_{k=0}^{\infty} L_k^{(n)}(x) \lambda^k =
(1- \lambda)^{-1-n}
\exp \frac{-x\lambda}{1-\lambda},~
L_k^{(n)}(x)=0~(k<0)\ .\label{Laguerre}
\end{equation}
Then we have the following theorem:
\renewcommand{\thesection}{\arabic{section}}
\begin{th}\label{main}
Let $\tau_N(n)$ be the $N\times N$ determinant given by
\begin{equation}
\tau_N(n)=\left|
\matrix{L_{N}^{(n)}    & L_{N+1}^{(n)}  & \cdots & L_{2N-1}^{(n)} \cr
	L_{N-2}^{(n)}  & L_{N-1}^{(n)}  & \cdots & L_{2N-3}^{(n)} \cr
	\vdots      & \vdots      & \ddots & \vdots      \cr
	L_{-N+2}^{(n)} & L_{-N+3}^{(n)} & \cdots & L_{1}^{(n)}
	}
\right|\ .\label{tau}
\end{equation}
Then 
\begin{equation}
X(n) = \frac{\tau_{N+1}(n+1)\tau_N(n-1)}{\tau_{N+1}(n)\tau_N(n)}-1\ ,\label{dep}
\end{equation}
satisfies dP$_{\rm II}$,
\begin{equation}
X(n+1) + X(n-1) = \frac{2}{x}\frac{(n+1)X(n)-(N+1)}{1-X(n)^2}\ .\label{eq}
\end{equation}
\end{th}
Note that $L_k^{(n)}(x)$ is the polynomial of $k$-th degree
in $n$, and hence eq.(\ref{dep}) yields the rational solution of
dP$_{\rm II}$.

Theorem \ref{main} follows directly from the following
proposition.

\begin{prop}\label{prop1}
The $\tau$ function (\ref{tau}) satisfies the following
bilinear equations:
\begin{equation}
\tau_{N+1}(n+1)\tau_{N}(n-1)+\tau_{N+1}(n-1)\tau_{N}(n+1)
	-2\tau_{N+1}(n)\tau_{N}(n)=0\ ,\label{bl1}
\end{equation}
\begin{equation}
\tau_{N}(n)\tau_{N}(n+1)-\tau_{N+1}(n+1)\tau_{N-1}(n)
	+\tau_{N+1}(n)\tau_{N-1}(n+1)=0\ ,\label{bl2}
\end{equation}
\begin{equation}
x\tau_{N}(n+2)\tau_{N}(n-1)
	-(n-N+1)\tau_{N}(n+1)\tau_{N}(n)
	+(2N+1)\tau_{N+1}(n)\tau_{N-1}(n+1)=0\ .\label{bl3}
\end{equation}
\end{prop}

Theorem 2.1 is derived from Proposition 2.2 as follows.
Introducing the variables by
\begin{eqnarray}
v_N(n)&=&\frac{\tau_{N+1}(n+1)}{\tau_{N}(n)}\ , \\
u_N(n)&=&\frac{\tau_{N}(n-2)\tau_{N}(n+1)}{\tau_{N}(n-1)\tau_{N}(n)}\ ,
\end{eqnarray}
then the bilinear equations (\ref{bl1})-(\ref{bl3}) are
rewritten as
\begin{equation}
v_N(n)+v_N(n-2)u_N(n)-2v_N(n-1)=0\ ,\label{e1}
\end{equation}
\begin{equation}
1+\frac{v_{N}(n-1)}{v_{N-1}(n+1)}u_N(n+1)
	-\frac{v_{N}(n)}{v_{N-1}(n)}=0\ ,\label{e2}
\end{equation}
\begin{equation}
xu_N(n+1)-(n-N+1)+(2N+1)\frac{v_N(n-1)}{v_{N-1}(n+1)}u_N(n+1)=0
\ ,\label{e3}
\end{equation}
respectively. Eliminating $v_{N-1}$ and $u_N$ from eqs.(\ref{e1})-(\ref{e3}),
and putting
\begin{equation}
X(n)=\frac{v_N(n)}{v_N(n-1)}-1\ ,
\end{equation}
we find that $X(n)$ satisfies eq.(\ref{eq}),
which is the desired result.

\section{Proof of Proposition 2.2}
In this section, we give the proof of Proposition \ref{prop1}.

Bilinear difference equations are derived from the Pl\"ucker relations
which are identities among determinants whose columns or rows
are shifted. By applying the Laplace expansion on certain determinants
which are identically zero, we obtain the Pl\"ucker relations.
We obtain the bilinear equations from the Pl\"ucker relations
with the aid of the difference formulas
that relate the ``shifted determinants'' with $\tau$ function
by using the contiguity relations of the entries,
\begin{equation}
L_k^{(n-1)}=L_k^{(n)}-L_{k-1}^{(n)}\ ,\label{rec1}
\end{equation}
\begin{equation}
kL_k^{(n)}
=(n+1)L_{k-1}^{(n+1)}-xL_{k-1}^{(n+2)}\ .\label{rec2}
\end{equation}

\subsection{Equation (8)}

First we prove that the $\tau$ function (\ref{tau}) satisfies
eq.(\ref{bl2}), which is the simplest example to demonstrate
the above procedure. 
We introduce a convenient notation:
\begin{equation}
\tau_N(n)=|-N+2_n, -N+3_n, \cdots, 1_n|\ ,
\end{equation}
where ``$j_n$'' denotes the column vector which ends with
$L_j^{(n)}$,
\begin{equation}
j_n = \pmatrix{\vdots\cr L_{j+2}^{(n)}\cr L_j^{(n)}\cr}\label{shift}
\end{equation}
Here the height of the column $j_n$ is $N$.
However in the following, we use the same symbol
for determinants with different size.
So the size of $j_n$ should be read appropriately
case by case. Moreover, we sometimes suppress the suffix $n$ if
there is no possibility of confusion.

Noticing that $L_0^{(n)}(x)=1$ and $L_k^{(n)}(x)=0$ for $k<0$, 
we see that the $\tau$ function is also expressible as
\begin{equation}
\tau_N(n)
=\left|
	\matrix{L_{N}^{(n)}    & L_{N+1}^{(n)}  &\cdots & L_{2N-1}^{(n)} & L_{2N}^{(n)} \cr
		L_{N-2}^{(n)}  & L_{N-1}^{(n)}  &\cdots & L_{2N-3}^{(n)} & L_{2N-2}^{(n)} \cr
		\vdots     & \vdots      & \ddots      &\vdots & \vdots      \cr
		L_{-N+2}^{(n)} & L_{-N+3}^{(n)} &\cdots & L_{1}^{(n)} & L_{2}^{(n)} \cr
		L_{-N}^{(n)}   & L_{-N+1}^{(n)} &\cdots & L_{-1}^{(n)} & L_{0}^{(n)}
	}
\right|,
\end{equation}
namely,
\begin{eqnarray}
\tau_N(n)
&=&\left|
	\matrix{-N+2 & -N+3 & \cdots & 1}
\right|\ ,\label{diff1z}
\\
&=&\left|
	\matrix{-N & -N+1 & \cdots & -1 & 0}
\right|\ .\label{diff1a}
\end{eqnarray}
Now subtracting $(i-1)$-th column from $i$-th column
of $\tau_N(n+1)$ for $i=N,\cdots 2$ and using eq.(\ref{rec1}), we have
\begin{eqnarray}
\tau_N(n+1)
&=&|\matrix{-N+2_{n+1} & -N+3 & \cdots & 1}|\ ,\label{diff1b}\\
&=&|\matrix{-N_{n+1} & -N+1 & \cdots & 0}|\ .\label{diff1c}
\end{eqnarray}
Moreover, adding the second column to the first column in eqs.(\ref{diff1b})
and (\ref{diff1c}), we get
\begin{eqnarray}
\tau_N(n+1)
&=&|\matrix{-N+3_{n+1} & -N+3 & \cdots & 1}|\ ,\label{diff1d}\\
&=&|\matrix{-N+1_{n+1} & -N+1 & \cdots & 0}|\ .\label{diff1e}
\end{eqnarray}
Equations (\ref{diff1z})-(\ref{diff1e}) are regarded
as ``difference formulas''.

Now consider the identity of $2N\times 2N$ determinant,
\begin{equation}
0=\left|
\matrix{
-N+2 &\vbl{4}&\matrix{ -N+2_{n+1} &  -N+3 & \cdots & 0}&\vbl{4} & \matrix{\hbox{\O}}&
\vbl{4}& 1 & \phi \cr
\multispan{8}\hblfil\cr
-N+2 &\vbl{4}&\matrix{\hbox{\O}}&\vbl{4}& \matrix{ -N+3 & \cdots & 0 }&\vbl{4}&
1 & \phi}\right|\ ,\label{pl1}
\end{equation}
where \hbox{\O}  means the empty block and
\begin{equation}
\phi = \pmatrix{1\cr 0\cr\vdots\cr 0}\ .
\end{equation}
Applying the Laplace expansion on the right hand side of eq.(\ref{pl1}),
we have
\begin{eqnarray}
0&=&\left|
\matrix{ -N+2 &-N+2_{n+1} & -N+3 & \cdots & 0}
\right|
\left|
\matrix{-N+3 & \cdots & 0 & 1 & \phi}
\right|
\nonumber \\
&+&\left|
\matrix{-N+2_{n+1} & -N+3 & \cdots & 0 & 1}
\right|
\left|
\matrix{-N+2 & -N+3 & \cdots & 0 & \phi}
\right|
\nonumber \\
&-&\left|
\matrix{-N+2_{n+1} & -N+3 & \cdots & 0 & \phi}
\right|
\left|
\matrix{-N+2 & -N+3 & \cdots & 0 & 1}
\right|\ .\nonumber\\
\end{eqnarray}
By using difference formulas (\ref{diff1z})-(\ref{diff1e}), we obtain,
\begin{equation}
0=-\tau_{N-1}(n+1)\tau_{N-1}(n) + \tau_N(n+1)\tau_{N-2}(n)
- \tau_{N-2}(n+1)\tau_N(n)\ ,
\end{equation}
which is equivalent to eq.(\ref{bl2}).

  \subsection{Equation (7)}
Secondly, we prove eq.(\ref{bl1}).
We introduce 
\begin{equation}
\ol{L}_{k}^{(n)}=L_{k}^{(n)}+L_{k-1}^{(n)},\label{bar}
\end{equation}
and
\begin{equation}
\ol{j}_n = \pmatrix{
\vdots\cr \ol{L}_{j+2}^{(n)}\cr \ol{L}_j^{(n)}\label{shift2}
}\ .
\end{equation}
Then we have the following difference formulas:
\begin{lem}\label{lemma1}
\begin{eqnarray}
 \tau_N(n+1)&=&|\ol{-N+2}_n, \ol{-N+3}_n, \cdots, \ol{1}_n|\ .\label{diff2a}\\
\tau_N(n+2)&=&|\matrix{\ol{-N+2_{n+1}} & \ol{-N+3} & \cdots & \ol{1}}|,
\label{diff2b}\\
\tau_N(n+2)&=&|\matrix{\ol{-N+3_{n+1}} & \ol{-N+3} & \cdots & \ol{1}}|,
\label{diff2c}\\
\tau_N(n)&=&|\matrix{-N+2 & \ol{-N+3} & \cdots & \ol{1}}|,\label{diff2d}\\
2\tau_N(n+1)&=&|\matrix{\ol{-N+3_{n+1}} & -N+3 & \ol{-N+4} & \cdots & \ol{1}}|,
\label{diff2e}\\
-\tau_N(n)&=&|\matrix{-N+3 & \ol{-N+3} & \cdots & \ol{1}}|\ .\label{diff2f}
\end{eqnarray}
Here, $j$ is the same as $j_n$ given in eq.(\ref{shift}).
\end{lem}
It is easy to prove lemma \ref{lemma1} by the similar
method to the derivation of eqs.(\ref{diff1z})-(\ref{diff1e}).
We give it in the appendix A.

Consider the following identity of $2N\times 2N$ determinant,
\begin{equation}
0=\left\vert\matrix{
\matrix{-N+3 & \ol{-N+3} } &\vbl{4} 
& \matrix{\ol{-N+3}_{n+1} & \ol{-N+4}&\cdots & \ol{1}}&\vbl{4} &\matrix{\hbox{\O}}
&\vbl{4} &\phi \cr
 \multispan{7}\hblfil \cr
\matrix{-N+3 & \ol{-N+3} } &\vbl{4} 
&\matrix{\hbox{\O}} &\vbl{4} &\matrix{\ol{-N+4}&\cdots & \ol{1}}
&\vbl{4} &\phi \cr
}\right\vert\ .
\end{equation}
Applying the Laplace expansion on the right hand side, we get
\begin{eqnarray}
0&=&
|-N+3,\ol{-N+3}_{n+1},\ol{-N+4}\cdots,\ol{1}|\times 
|\ol{-N+3},\ol{-N+4},\cdots,\ol{1},\phi|
\nonumber\\
&-&|\ol{-N+3},\ol{-N+3}_{n+1},\ol{-N+4},\cdots,\ol{1}|\times 
|-N+3,\ol{-N+4},\cdots,\ol{1},\phi|\nonumber\\
&-&|\ol{-N+3}_{n+1},\ol{-N+4},\cdots,\ol{1},\phi|\times
 |-N+3,\ol{-N+3},\ol{-N+4},\cdots,\ol{1}|\nonumber\\
\end{eqnarray}
Then we obtain by using lemma\ref{lemma1},
\begin{equation}
0=-2\tau_{N}(n+1)\tau_{N-1}(n+1)
 - (-\tau_N(n+2))\cdot\tau_{N-1}(n)
-\tau_{N-1}(n+2)\cdot (-\tau_N(n))\ ,
\end{equation}
which is equivalent to eq.(\ref{bl1}). 

  \subsection{Equation (9)}
Finally, we prove eq.(\ref{bl3}).
We rewrite the $\tau$ function (\ref{tau})
as follows. Subtracting  $(i+1)$-th column  from $i$-th column for
$i=1,\cdots,j$, $j=N-1,\cdots 1$, and using eq.(\ref{rec1}) we get 
\begin{eqnarray}
\tau_N(n)
&=&(-1)^{\frac{N(N-1)}{2}}\left|
	\matrix{L_{2N-1}^{(n-N+1)} & L_{2N-1}^{(n-N+2)} & \cdots & L_{2N-1}^{(n)} \cr
		L_{2N-3}^{(n-N+1)} & L_{2N-3}^{(n-N+2)} & \cdots & L_{2N-3}^{(n)} \cr
		\vdots             & \vdots             & \ddots & \vdots      \cr
		L_{1}^{(n-N+1)}    & L_{1}^{(n-N+2)}    & \cdots & L_1^{(n)}
	}
\right|
\nonumber \\
&=&\left|
	\matrix{L_{2N-1}^{(n)} & L_{2N-1}^{(n-1)} & \cdots & L_{2N-1}^{(n-N+1)} \cr
		L_{2N-3}^{(n)} & L_{2N-3}^{(n-1)} & \cdots & L_{2N-3}^{(n-N+1)} \cr
		\vdots         & \vdots           & \ddots & \vdots \cr
		L_1^{(n)}      & L_1^{(n-1)}      & \cdots & L_1^{(n-N+1)}
	}
\right|\ .\label{tau-n}
\end{eqnarray}
Here, we introduce a notation,
\begin{equation}
\tau_N(n)=|[n][n-1]\cdots [n-N+1]|\ ,
\end{equation}
where ``$[j]$'' denotes the column vector,
\begin{equation}
[j]= \pmatrix{\vdots\cr ~L_{3}^{(j)}\cr ~L_{1}^{(j)}\cr}\ .\label{bracket}
\end{equation}
Then we have the following difference formulas:
\begin{lem}\label{lemma3}
\begin{equation}
 \frac{(-x)^N}{{\displaystyle \prod_{j=1}^N(2j+1)}}\tau_N(n)=
		|[n-2]\cdots [n-N-1] \wh{[n-N]}|\ ,\label{diff3a}
\end{equation}
\begin{equation}
\frac{-(n-N-1)(-x)^{N-1}}{{\displaystyle \prod_{j=1}^N(2j+1)}}\tau_N(n-1)=
	|[n-3] \cdots [n-N-2] \wh{[n-N]}|\ ,\label{diff3b}
\end{equation}
where $[\hat j]$ is the column vector,
\begin{equation}
[\hat j]= \pmatrix{\vdots \cr \hat L_2^{(j)}\cr \hat L_0^{(j)}}\ ,
\label{hat_bracket}
\end{equation}
and
\begin{equation}
\hat L_k^{(n)} = \frac{L_k^{(n)}}{k+1}\ .\label{hat}
\end{equation}
\end{lem}
We give the proof of lemma \ref{lemma3} in appendix B.

By applying the Laplace expansion on the following determinant which is
identically zero,
\begin{eqnarray*}
0&=&
\left|
\matrix{\matrix{[n-2]\!\!\!\!\! & [n-3]\!\!\!\!\! & \cdots\!\!\!\!\! &[n-N-1]} &\vbl{4} & \matrix{\hbox{\O}}
&\vbl{4} &\matrix{[n-N-2]\!\!\!\!\! & \wh{[n-N]} & \phi} \cr
\multispan{5}\hblfil\cr
 \matrix{\hbox{\O}}&\vbl{4} &\matrix{[n-3]\!\!\!\!\! & \cdots\!\!\!\!\! & [n-N-1]}&\vbl{4} &
\matrix{[n-N-2]\!\!\!\!\! & \wh{[n-N]} & \phi} }
\right|
\end{eqnarray*}
we get,
\begin{eqnarray*}
&&|\;[n-2]\; [n-3]\; \cdots [n-N-1]\; [n-N-2]\;|\\
&&\times|\;[n-3]\; \cdots [n-N-1]\; \wh{[n-N]}\; \phi\;|\\
&-&|\;[n-2]\; [n-3]\; \cdots [n-N-1]\; \wh{[n-N]}\;|\\
&&\times|\;[n-3]\; \cdots [n-N-1]\; [n-N-2]\; \phi\;|\\
&+&|\;[n-2]\; [n-3]\; \cdots [n-N-1]\; \phi\;|\\
&&\times|\;[n-3]\; \cdots [n-N-1]\; [n-N-2]\; \wh{[n-N]}\;|=0\ .
\end{eqnarray*}
Then we obtain by using lemma\ref{lemma3},
\begin{eqnarray*}
&&\tau_{N+1}(n-2)\frac{(-x)^{N-1}}{{\displaystyle\prod_{j=1}^{N-1}(2j+1)}}
\tau_{N-1}(n-1)
-\frac{(-x)^N}{{\displaystyle\prod_{j=1}^N(2j+1)}}\tau_N(n)\tau_N(n-3)\\
&&+\tau_N(n-2)\frac{-(n-N-1)(-x)^{N-1}}{{\displaystyle\prod_{j=1}^N(2j+1)}}
\tau_N(n-1)
=0\ , 
\end{eqnarray*}
which yields eq.(\ref{bl3}).
Thus we have proved Proposition 2.2, and hence Theorem 2.1.

\section{Continuous Limit}

In this section, we consider the continuous limit of the rational
solutions of dP$_{\rm II}$ to those of 
P$_{\rm II}$(\ref{p2}). 

In ref.\cite{P2}, it is shown that the rational solutions
for P$_{\rm II}$(\ref{PII})
are given as follows:
\begin{prop}
Let $p_k(z,t)$ be the Devisme polynomial defined by
\begin{equation}
\sum_{k=0}^\infty p_k(z,t)\eta^k = \exp\left(z\eta+t\eta^2+\frac{1}{3}\eta^3
\right)\ ,\quad p_k(z,t)=0\ {\rm for}\ k<0\ \label{Devisme}
\end{equation}
and let $\tau_N$ be an $N\times N$ determinant given by
\begin{equation}
\tau_N=\left|\matrix{
p_{N}(z,t) & p_{N+1}(z,t) & \cdots & p_{2N-1}(z,t)\cr
p_{N-2}(z,t) & p_{N-1}(z,t) & \cdots & p_{2N-3}(z,t)\cr
\vdots  & \vdots    & \ddots & \vdots \cr
p_{-N+2}(z,t) & p_{-N+3}(z,t) & \cdots & p_{1}(z,t)\cr}\right|\ ,
 \label{P2tau}
\end{equation}
Then 
\begin{equation}
v=\frac{d}{dz}\log\frac{\tau_{N+1}}{\tau_{N}},\label{var}
\end{equation}
satisfies P$_{\rm II}$,
\begin{equation}
\frac{d^2}{dz^2}v = 2v^3 - 4zv + 4(N+1)\ .
\label{PII}
\end{equation}
\end{prop}
\begin{rem}
The $\tau$ function (\ref{P2tau}) does not depend on $t$.
\end{rem}

Let us consider the continuous limit of the result in Theorem \ref{main}.
We shift $n$ in eq.(\ref{eq}) by $x-1$ 
\begin{equation}
X^\prime(n+1) + X^\prime(n-1) = \frac{2}{x}\frac{(n+x)X^\prime(n)-(N+1)}{1-X^\prime(n)^2}\ ,\label{eqs}
\end{equation}
where $X^\prime(n)=X(n+x-1)$.
Putting 
\begin{equation}
x=-\frac{1}{2\varepsilon^3},\quad n=\frac{z}{\varepsilon},\quad
X^\prime(n) = \varepsilon v\ ,\label{par}
\end{equation}
then we easily find that eqs.(\ref{dep}) and (\ref{eqs}) reduce
to eqs.(\ref{var}) and (\ref{PII}), respectively, in the limit
of $\varepsilon\rightarrow 0$. 

Then, how about the solutions? Unfortunately, $L_k^{(n+x-1)}(x)$,
which is the entry of the $\tau_N(n+x-1)$, 
does not reduce to $p_k$ (\ref{Devisme}) in this limit.
We need some trick to adjust them as described in the
following proposition.
\begin{prop}\label{con}
Let $\hat L_k^{(n)}(x)$ be the polynomial defined by
\begin{equation}
\sum_{k=0}^{\infty} \hat{L}_k^{(n)}(x) \lambda^k
=(1-\lambda^2)^{-\frac{1}{2}x}(1- \lambda)^{-n-x}
\exp \frac{-x\lambda}{1-\lambda},\quad\hat{L}_k^{(n)}(x)=0~(k<0)\ .
\label{lag2}
\end{equation}
Then $\tau_N(n+x-1)$ is expressed as
\begin{equation}
\tau_N(n+x-1)=\left|
\matrix{\hat L_{N}^{(n)}    & \hat L_{N+1}^{(n)}  & \cdots & \hat L_{2N-1}^{(n)} \cr
	\hat L_{N-2}^{(n)}  & \hat L_{N-1}^{(n)}  & \cdots & \hat L_{2N-3}^{(n)} \cr
	\vdots      & \vdots      & \ddots & \vdots      \cr
	\hat L_{-N+2}^{(n)} & \hat L_{-N+3}^{(n)} & \cdots & \hat L_{1}^{(n)}
	}
\right|\ ,
\end{equation}
and $\varepsilon^k\hat L_k^{(n)}(x)$ reduces to $p_k(z,0)$ in the limit
of $\varepsilon\rightarrow 0$ with the parametrization (\ref{par}).
Moreover, $X^\prime(n)$ reduces to rational solutions of P$_{\rm II}$
in this limit.
\end{prop}
{\it Proof of Proposition \ref{con}}:\\
The first statement is verified easily, since 
$\hat L^{(n)}_k(x)$ is a linear combination of $L^{(n+x-1)}_j(x)$, 
$j=k,k-2,k-4,\cdots$. The second statement is also checked
by putting $\lambda=\varepsilon\eta$ in eq.(\ref{lag2}),
choosing the parameters as eq.(\ref{par}) and taking the
limit of $\varepsilon\rightarrow 0$. Although
multiplication of $\varepsilon^k$ on $\hat L_k$ yields overall factor 
$\varepsilon^{N(N+1)/2}$ on $\tau_N(n+x-1)$, this does not
make any effect on $X^\prime(n)$. Hence the rational solutions of
dP$_{\rm II}$ reduces to those of P$_{\rm II}$ in the continuous
limit. \hfill\qed

Finally, we mention on the continuous limit of the bilinear
equations. It is shown that the $\tau$ function of P$_{\rm II}$
(\ref{P2tau}) satisfies
\begin{eqnarray}
&&D_z^2 g\cdot f = 0,
\label{P2bl1}\\
&&\left( D_z^3 + 4zD_z - 4(N+1)\right) g\cdot f = 0\ ,
\label{P2bl2}
\end{eqnarray}
where $D_z^n$ is Hirota's bilinear operator defined by
\begin{equation}
D_z^ng\cdot f =\left. \left(\partial_z-\partial_{z^\prime}\right)^n
g(z)f(z^\prime)\right|_{z=z^\prime}\ .
\end{equation}
Combining three bilinear difference equations (\ref{bl1})--(\ref{bl3})
for dP$_{\rm II}$, we can show that $\tau$ function of dP$_{\rm II}$
(\ref{tau}) satisfies\cite{dP2_rational}
\begin{equation}
\left[\cosh D_n-1\right]~\tau_{N+1}(n)\cdot \tau_N(n)=0\ ,\label{bl4}
\end{equation}
\begin{equation}
\left[ -\frac{x}{2}\sinh 2D_n
	+( n+1 )\sinh D_n
	-(N+1)\right] \tau_{N+1}(n)\cdot \tau_N(n)=0\ ,\label{bl5}
\end{equation}
where 
\[
\cosh D_n = \frac{{\rm e}^{D_n}+{\rm e}^{-D_n}}{2},\quad
\sinh D_n = \frac{{\rm e}^{D_n}-{\rm e}^{-D_n}}{2},
\]
and 
\[
 {\rm e}^{D_n}f(n)\cdot g(n)=f(n+1)g(n-1)\ .
\]
Then, after replacing $n$ by $n+x-1$,
it is easy to see that equations (\ref{bl4}) and (\ref{bl5})
reduce to eqs.(\ref{P2bl1}) and (\ref{P2bl2}) respectively,
in the limit $\varepsilon\rightarrow 0$ with the parametrization (\ref{par}).

\section{Conclusion}
In this article, we have constructed the rational solutions
for dP$_{\rm II}$, and shown that they are expressed
by the determinants whose entries are given by
the Laguerre polynomials.
We have also discussed the continuous limit to P$_{\rm II}$.

It is known that the Laguerre polynomials appear
in the rational solutions for continuous P$_{\rm V}$\cite{PV}.
Moreover, the discrete Airy functions which appeared in the
special function type solutions for dP$_{\rm II}$ can be
expressed in terms of the Hermite-Weber functions\cite{dP2,dP1},
which are also the solutions of P$_{\rm IV}$\cite{Okamoto}.
Moreover, so-called the ``molecular type solution'' of dP$_{\rm II}$
is expressed by the same $\tau$ function as that of the Bessel function
type solution for P$_{\rm III}$\cite{Kokyuroku}.
At present, we do not know what these strange relations mean.

There are so many
versions of discrete Painlev\'e equations, namely, there are many
difference equations which passes the singularity confinement test\cite{SC}
and reduces to the same Painlev\'e equation in the continuous limit.
However, their solutions
have been constructed only for a few cases. It may be
an important problem to find those solutions, since explicit forms
of the particular solutions would be one of the most
useful keys to understand the discrete Painlev\'e equations.

The authors would like to thank Profs. B.Grammaticos, A.Ramani and
J. Hietarinta for encouragement and discussions. 
One of the authors(K.K) was supported by the Grant-in-Aid for Encouragement
of Young Scientists from The Ministry of Education, Science,
Sports and Culture of Japan, No.08750090.

\appendix
\section{Proof of Lemma 3.1}

We have from (\ref{rec1}) and (\ref{bar}),
\begin{equation}
L_{k}^{(n)}-L_{k-2}^{(n)}=\ol{L}_{k}^{(n-1)}\ ,\label{rec4}
\end{equation}
\begin{equation}
2L_{k}^{(n)}=L_{k}^{(n-1)}+\ol{L}_{k}^{(n)}\ .\label{rec5}
\end{equation}
In eq.(\ref{tau}), subtracting $(i+1)$-th row
from $i$-th row for $i=1,\cdots N-1$, we have
\begin{eqnarray}
\tau_N(n)
&=&
\left|
	\matrix{\ol{L}_N^{(n-1)}    & \ol{L}_{N+1}^{(n-1)}  & \cdots & \ol{L}_{2N-1}^{(n-1)} \cr
		\ol{L}_{N-2}^{(n-1)}  & \ol{L}_{N-1}^{(n-1)}  & \cdots & \ol{L}_{2N-3}^{(n-1)} \cr
		\vdots              & \vdots              & \ddots & \vdots      \cr
		\ol{L}_{-N+2}^{(n-1)} & \ol{L}_{-N+3}^{(n-1)} & \cdots & \ol{L}_1^{(n-1)}
	}
\right|,
\end{eqnarray}
where, for $N$-th row, we used $L_{k}^{(n)}=\ol{L}_{k}^{(n-1)}$.
Thus we get eq.(\ref{diff2a}).
Similarly, from eqs.(\ref{diff1b}) and (\ref{diff1d}), 
we obtain eqs.(\ref{diff2b}) and (\ref{diff2c}), respevtively.

Next, adding $(i-1)$-th column to $i$-th column of eq.(\ref{tau})
for $i=N,\cdots,2$ and using eq.(\ref{bar}), we get
eq.(\ref{diff2d}).
Similarly, adding $(i-1)$-th column to $i$-th column of eq.(\ref{diff1d})
for $i=N,\cdots,3$, we get
\begin{equation}
\tau_N(n+1)=|\matrix{-N+3_{n+1} & -N+3 & \ol{-N+4} & \cdots & \ol{1}}|.
\end{equation}
Now we obtain by using eq.(\ref{rec5}),
\begin{eqnarray}
2\tau_N(n+1)&=&|\matrix{2(-N+3_{n+1}) & -N+3 & \ol{-N+4} & \cdots & \ol{1}}| \nonumber \\
&=&|\matrix{-N+3 & -N+3 & \ol{-N+4} & \cdots & \ol{1}}| \nonumber \\
&& \quad +|\matrix{\ol{-N+3_{n+1}} & -N+3 & \ol{-N+4} & \cdots & \ol{1}}| \nonumber \\
&=&|\matrix{\ol{-N+3_{n+1}} & -N+3 & \ol{-N+4} & \cdots & \ol{1}}|, 
\end{eqnarray}
which is nothing but eq.(\ref{diff2e}).
Finally, eq.(\ref{diff2f}) is derived by 
subtracting the second column from the first column
in eq.(\ref{diff2d}).

\section{Proof of Lemma 3.2}

First, notice that the $\tau$ function can also be written as
\begin{eqnarray}
\tau_N(n)
&=&\left|
	\matrix{L_N^{(n)}     & L_{N+1}^{(n)}  & \cdots & L_{2N}^{(n)} \cr
		L_{N-2}^{(n)} & L_{N-1}^{(n)}  & \cdots & L_{2N-2}^{(n)} \cr
		\vdots        & \vdots         & \ddots & \vdots      \cr
		L_{-N}^{(n)}  & L_{-N+1}^{(n)} & \cdots & L_0^{(n)}
	}
\right|
\nonumber \\
&=&\left|
	\matrix{L_{2N}^{(n)}   & L_{2N}^{(n-1)}   & \cdots & L_{2N}^{(n-N)} \cr
		L_{2N-2}^{(n)} & L_{2N-2}^{(n-1)} & \cdots & L_{2N-2}^{(n-N)} \cr
		\vdots         & \vdots           & \ddots & \vdots \cr
		L_0^{(n)}      & L_0^{(n-1)}      & \cdots & L_0^{(n-N)}
	}
\right|\ .
\label{tau-n1}
\end{eqnarray}
Now adding $(i+1)$-th column multiplied by $(n-i)/(-x)$ 
to $i$-th column of eq.(\ref{tau-n1}) 
for $i=1,\cdots,N$, and using eq.(\ref{rec2}), we have
\begin{eqnarray}
\tau_N(n)
&=&\frac{1}{(-x)^N}
\left|
	\matrix{(2N+1)L_{2N+1}^{(n-2)} & \cdots & (2N+1)L_{2N+1}^{(n-N-1)} & L_{2N}^{(n-N)} \cr
		(2N-1)L_{2N-1}^{(n-2)} & \cdots & (2N-1)L_{2N-1}^{(n-N-1)} & L_{2N-2}^{(n-N)} \cr
		\vdots                 & \ddots & \vdots  & \vdots \cr
		L_1^{(n-2)}            & \cdots & L_1^{(n-N-1)}         & L_0^{(n-N)}
	}
\right|
\label{shift3a} \\
&=&\frac{{\displaystyle\prod_{j=1}^N(2j+1)}}{(-x)^N}
\left|
	\matrix{L_{2N+1}^{(n-2)} & \cdots & L_{2N+1}^{(n-N-1)} & {\hat L}_{2N}^{(n-N)} \cr
		L_{2N-1}^{(n-2)} & \cdots & L_{2N-1}^{(n-N-1)} & {\hat L}_{2N-2}^{(n-N)} \cr
		\vdots           & \ddots & \vdots             & \vdots \cr
		L_{1}^{(n-2)}    & \cdots & L_{1}^{(n-N-1)}    & {\hat L}_{0}^{(n-N)}
	}
\right| ,\label{shift3b}
\end{eqnarray}
where ${\hat L}_{k}^{(n)}$ is given in eq.(\ref{hat}). 
This gives eq.(\ref{diff3a}).

Moreover, adding $N$-th column to $(N+1)$-th column multiplied by $-(n-N)$
of eq.(\ref{shift3a}), we get
\begin{eqnarray}
\tau_N(n)
&=&\frac{1}{-(n-N)(-x)^{N-1}}
\left|
	\matrix{(2N+1)L_{2N+1}^{(n-2)} & \cdots & (2N+1)L_{2N+1}^{(n-N-1)} & L_{2N}^{(n-N+1)} \cr
		(2N-1)L_{2N-1}^{(n-2)} & \cdots & (2N-1)L_{2N-1}^{(n-N-1)} & L_{2N-2}^{(n-N+1)} \cr
		\vdots                 & \ddots & \vdots  & \vdots \cr
		L_1^{(n-2)}            & \cdots & L_1^{(n-N-1)}         & L_0^{(n-N+1)}
	}
\right| \\
&=&\frac{{\displaystyle\prod_{j=1}^N(2j+1)}}{-(n-N)(-x)^{N-1}}
\left|
	\matrix{L_{2N+1}^{(n-2)} & \cdots & L_{2N+1}^{(n-N-1)} & {\hat L}_{2N}^{(n-N+1)} \cr
		L_{2N-1}^{(n-2)} & \cdots & L_{2N-1}^{(n-N-1)} & {\hat L}_{2N-2}^{(n-N+1)} \cr
		\vdots           & \ddots & \vdots             & \vdots \cr
		L_{1}^{(n-2)}    & \cdots & L_{1}^{(n-N-1)}    & {\hat L}_{0}^{(n-N+1)}
	}
\right|\ .\label{shift3c}
\end{eqnarray}
which gives eq.(\ref{diff3b}).

\end{document}